# Human-on-the-loop Resilient Control of Inverter-Based Resources Under Actuator Degradation

Majid Dehghani, Taha Saeed Khan, Student *Member, IEEE*, Hamidreza Nazaripouya, *Senior Member, IEEE*

***Abstract*—This paper proposes a human-on-the-loop (HOTL) resilient control architecture for grid-supporting inverter-based resources (IBRs) operating under actuator degradation. Conventional fault-tolerant control (FTC) and adaptive control strategies each face notable limitations in this setting: active FTC depends on fast, accurate fault detection and isolation, leaving it vulnerable to misdiagnosis of incipient or ambiguous degradation, while passive FTC tends toward overly conservative operation. Adaptive controllers face a related problem as they typically assume sufficient control authority, but when actuator degradation erodes that authority, the adaptive law may misinterpret tracking errors, leading to parameter drift, performance loss, or instability. To overcome these limitations, the proposed framework embeds human supervisory judgment directly into the control loop, detecting subtle off-nominal behavior, validating or overriding controller parameters, and adjusting operational objectives when conditions exceed the modeled fault space. Two new metrics underpin this resilient decision-making: Generation Reserve Capacity (GRC), which quantifies remaining inverter capacity available for future contingencies, and Controlled Performance Degradation (CPD), which allows temporary, deliberate performance relaxation to preserve overall system operability. A human-selected resilience tuning parameter, $\mu$, governs the trade-off between immediate tracking accuracy and long-term operational readiness. A stability analysis of the proposed HOTL $\mu$-mod adaptive controller is presented. The approach is also validated through simulations of a grid-connected inverter under sequential actuator degradation. Results show that the proposed architecture preserves control reserves, prevents actuator saturation, and achieves superior voltage regulation compared with conventional adaptive control and FTC methods.**

## I. INTRODUCTION

Modern power distribution networks are undergoing a rapid shift driven by the widespread deployment of inverter-based resources (IBRs) [1], [2], including solar photovoltaic (PV) systems [3], fuel cells, type 3 and 4 wind turbines, and battery energy storage systems [4]. As these IBRs are increasingly growing [5], distribution feeders are becoming dominated by fast-response power electronic inverters. These IBRs provide essential services—such as voltage regulation [6] and reactive-power support [7]. However, their operational performance may be substantially degraded under contingencies, where even moderate IBR degradation can significantly impair voltage support, limit reactive power capability, and compromise the distribution system's ability to maintain secure and stable operation during subsequent disturbances.

Contingencies affecting IBRs can be broadly categorized as communication/cyber-related [8], [9], [10], and actuator/physical-related [11], [12]. To ensure resilient operation under these contingencies, resilient control has emerged as a key functional requirement for IBRs [13], [14], [15], [16]. Accordingly, significant research efforts have been devoted to developing resilient control strategies for inverter-based power systems [17], [18], [19], [20]. Existing resilient-control approaches can also be classified into two categories: communication/cyber-resilient control and actuator/physical-resilient control.

Communication/cyber-resilient control methods focus on maintaining system performance under corrupted measurements and communication-layer contingencies. Existing studies have employed adaptive, observer-based, and robust control strategies to mitigate cyber-attacks, communication contingencies, and measurement corruption in IBR-dominated systems. Ref [21] proposes a resilient control for frequency regulation in IBRs that remains robust under cyber-attacks using an adaptive Type-2 fuzzy PID controller. Similarly, Ref [22] proposes an adaptive resilient secondary control scheme for islanded AC microgrids that compensates unknown, time-varying measurement corruption without requiring prior knowledge of attack bounds. Ref [23] proposes a resilient $H_\infty$ control method for nonlinear systems with bounded uncertainties and external disturbances, including sensor and communication attacks, without relying on online fault estimation. In Ref [24] a hybrid-observer-based resilient control strategy is developed to mitigate delayed and sporadic measurement and communication attacks in load–frequency regulation systems. Ref [25] proposes a distributed integral control strategy for power systems with synchronous generators and IBRs to maintain stability under contingencies such as cyber-attacks and measurement faults using invariant-set theory and neighbor-to-neighbor communication.

The second category, including actuator/physical-resilient

Majid Dehghani is with Electrical Engineering Department, Oklahoma State University, Stillwater, OK, USA (e-mail: majid.dehghani@ okstate.edu).

Taha Saeed Khan is with Electrical Engineering Department, Oklahoma State University, Stillwater, OK, USA. (e-mail: taha_saeed.khan@okstate.edu).

Hamidreza Nazaripouya is with the Electrical Engineering Department, Oklahoma State University, Stillwater, OK, USA (e-mail: hanazar@okstate.edu).

control methods, addresses degradation in physical components and actuator. The resilient control method in [26] primarily focuses on maintaining system stability under actuator degradation. Ref [27] proposes a distributed active power control framework to address active power sharing and current control in large-scale IBR systems under grid voltage variations and partial or complete loss of inverters. Ref [28] introduces a distributed resilient secondary control framework for islanded microgrids that compensate for actuator faults through local adaptive compensators, restoring voltage and frequency deviations induced by droop control.

While both categories have advanced the state of resilient control for IBRs, several important limitations remain. First, existing approaches in both categories treat resilience enhancement as a fully automated process and do not incorporate real-time adaptation based on system operator expertise and situational awareness. These methods rely solely on model-based mechanisms — such as fault detection and isolation, observer-based estimation, or invariant-set theory — that are inherently limited by their predefined fault sets and modeling assumptions. As a result, they lack the contextual judgment needed to recognize subtle off-nominal behavior, validate or override control parameters, and adjust mission goals or trajectories when degradation falls outside the modeled fault set.

Second, while frameworks such as [26]- [28] compensate for degradation in individual actuators or inverter units, they do not provide a systematic, tunable mechanism for balancing the preservation of inverter reserve capacity against control performance during degraded operation. These approaches apply fixed compensation strategies without quantifiable resilience metrics or a means to adjust the trade-off between reserve margin and tracking accuracy based on the severity of the contingency.

These limitations highlight the need for a unified control framework that can simultaneously incorporate human supervisory intelligence, address actuator degradation, maintain reserve capacity, and provide rigorous stability guarantees for the closed-loop system under degraded operating conditions.

Motivated by these challenges, this paper introduces a human-on-the-loop (HOTL) adaptive control architecture designed to enhance the resilient operation of IBRs under actuator degradation, caused by cyber-physical attacks or physical faults. The proposed framework integrates a fast nonlinear adaptive control layer with a supervisory human decision-making layer, enabling the controller to incorporate real-time operator situational awareness and expertise into the control process. To support this coordination, two resilience metrics are developed: Generation Reserve Capacity (GRC), which quantifies the available inverter capacity margin and reflects the system's readiness for future contingencies; and Controlled Performance Degradation (CPD), which provides a principled mechanism for temporarily relaxing voltage-tracking performance during contingencies. Using these metrics, operators select an allowable CPD level through a resilience tuning parameter, while the adaptive controller rapidly adjusts the exchanged active and reactive power to maintain system states such as bus voltages within permissible operational bounds. Through this coordinated human–automation interaction, the proposed framework preserves inverter reserve capacity, prevents the control effort from reaching physical actuator saturation limits, and improves system resilience compared with conventional fault-tolerant and adaptive control approaches.

The main contributions of this paper are summarized as follows:

- Proposing a novel human-on-the-loop resilient control architecture that integrates operator situational awareness with adaptive inverter controller, enabling coordinated decision-making under actuator degradation and operating contingencies. In this framework, a supervisory mechanism is proposed for operator-guided controller tuning, where human operators evaluate the inverter-operating conditions and select the resilience parameter $\mu$ to balance immediate tracking performance against future operational readiness.
- Introducing two quantitative resilience metrics—GRC and CPD—that characterize the relationship between inverter reserve capacity, performance, and system survivability during contingencies.
- Proposing a $\mu$-mod adaptive control law with dynamically adjustable reference-model behavior, allowing the inverter to operate under partial capacity degradation while ensuring stable, graceful deviation from nominal tracking objectives.
- Providing a rigorous stability analysis of the proposed HOTL adaptive control architecture, proving that under the $\mu$-mod adaptive control law with dynamically adjustable reference-model behavior, all closed-loop signals remain bounded and the state tracking error converges asymptotically, even under actuator degradation caused by cyber-physical attacks or faults

The remainder of this paper is organized as follows. Section II introduces the problem formulation, including the mathematical model of the system, the formulation of inverter degradation and contingency conditions. Section III presents the proposed HOTL $\mu$-mod adaptive control architecture, describing the adaptive control layer, the supervisory human layer, and the resilience metrics that coordinate their interaction. Section IV introduces two conventional approaches—fault-tolerant control and baseline adaptive control—for comparison with the proposed controller. Section V demonstrates the effectiveness of the proposed method through simulation studies on a grid-connected inverter subjected to sequential degradation events. Finally, Section VI provides conclusion remarks.

## II. Problem Formulation

In inverter-supported distribution systems, inverter-based resources (IBRs) serve as actuators that implement control actions determined by supervisory or local controllers. Based on system measurements and control objectives, these controllers generate signals that are executed by IBRs through fast power-electronic switching and embedded control loops. In this role, IBRs act as the interface between control decisions and the physical power network, enabling direct and flexible manipulation of system variables in modern distribution systems.

This section formulates the dynamic model of an active distribution network with IBRs, then defines the control objectives under contingency conditions. The goal is to develop a HOTL control architecture that maintains resilient operation of IBRs through coordinated interaction between the human operator and the adaptive controller. The proposed framework aims to ensure stable voltage regulation and power dispatch under contingencies, cyber-physical attacks, and degraded operating conditions.

### *A. Distribution Network Model*

The generic model of an active distribution system including IBRs and composite load models is represented as (1) [29]:

$$\dot{x}(t) = Ax(t) + B\Lambda_f u(t) + d + \Phi^T f(x); \quad y(t) = Cx(t) \tag{1}$$

where $x(t) \in R^n$ is the state of the system, such as bus voltages, from steady state operating point, $u(t) \in R^m$ represents the control input vector, including the active and reactive power references of IBRs. The term $d$ represents uncertainties associated with the steady state condition, and $\Phi^T f(x)$ represents higher order effects due to nonlinearities with $\Phi$ as an unknown vector parameter.

The matrix $\Lambda_f = \text{diag}(\lambda_{f_1}, \lambda_{f_2}, \ldots, \lambda_{f_m})$ models the impact of contingencies, where each $\lambda_{f_i} \in (0,1]$ denotes the fraction of the total capacity of $IBR_i$ that remains effectively available under adverse conditions such as cyber-attacks or varying weather conditions. When no degradation is present, $\Lambda_f = I_m$.

To reflect realistic inverter operating conditions, the control inputs are constrained by physical power limits of inverters as:

$$u\,(t) = u_{max_i} \cdot sat(\frac{u_c(t)}{u_{max}}) = \begin{cases} u_c\,(t), & |\, u_c\,(t)\,| \leq u_{max} \\ u_{max_i} \cdot sgn(u_c\,(t)) & |\, u_c\,(t)\,| > u_{max} \end{cases} \tag{2}$$

where $u_{max_i}$ denotes the maximum allowable active or reactive power of $IBR$ and $u_c\,(t)$ is the control signal generated by the controller. The control objective is to regulate the distribution system states $x(t)$ (e.g., bus voltages) so that they follow a desired reference trajectory that is adaptively adjusted under system contingencies while remaining within permissible operational bounds.

### *B. Control Objectives and Resilience Concepts*

The proposed HOTL control architecture integrates an adaptive controller layer with cognitive engineering principles to incorporate the human operator's role in maintaining power distribution-system resilience. The framework introduces two resilience metrics, namely generation reserve capacity (GRC) and controlled performance degradation (CPD)—that guide control decisions during contingencies:

- **Generation Reserve Capacity (GRC):** GRC denotes the portion of system capacity that remains available for future contingencies. It quantifies the extra power capacity of IBRs beyond current demand and thus represents the network's reserve capacity for responding to future events. Maintaining sufficient GRC is critical for avoiding IBR saturation and ensuring resilient operation.
- **Controlled Performance Degradation (CPD):** CPD represents the tolerable reduction in performance (e.g., voltage tracking) post-contingency. When a severe contingency occurs, exact tracking of pre-contingency references may not be feasible. CPD acts as a tunable variable—adjusted by the operator—to temporarily relax performance requirements, thereby preserving GRC and overall system stability.

The primary goal of the control system is to maximize GRC with minimal CPD, maintaining states (e.g. voltages) within permissible operational bounds. During operation, when a contingency occurs, the human operator is tasked with high level cognitive tasks, such as perception of contingencies and an estimate of allowable CPD to achieve maximum GRC. The distributed energy resource management systems (DERMS), which are equipped with an adaptive controller, then rapidly dispatches IBRs to ensure stable system behavior. Fig. 1 depicts the proposed HOTL adaptive control architecture.

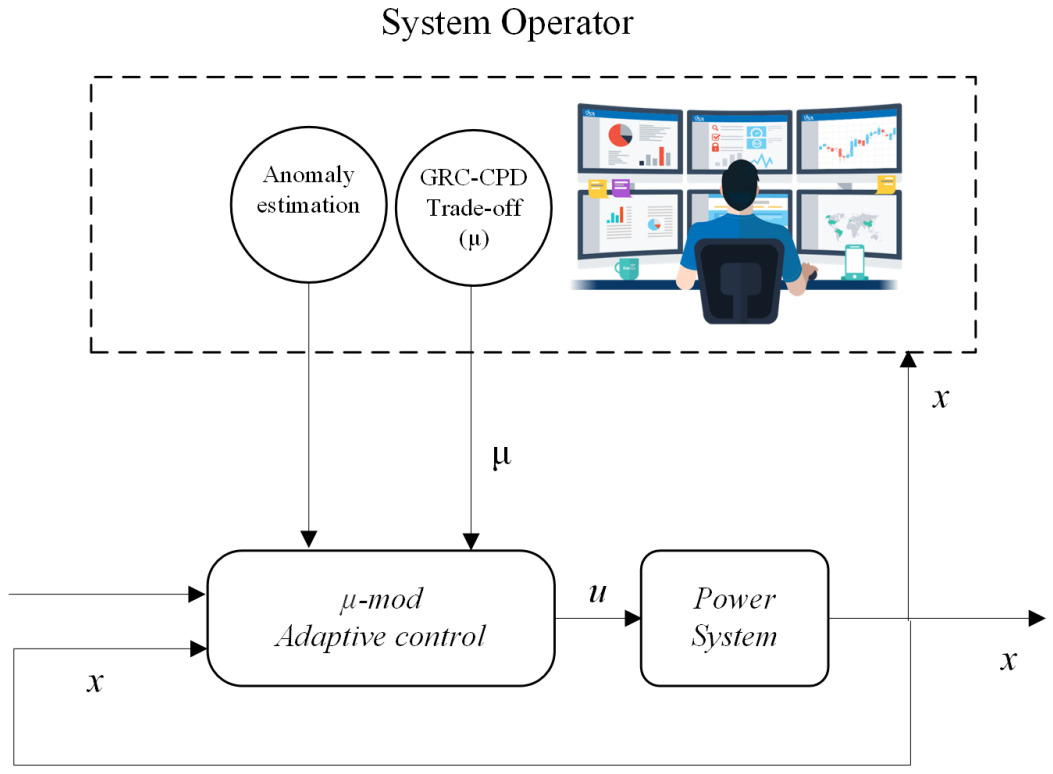

**Fig. 1.** HOTL Control Architecture.

### *C. HOTL Control Problem*

The overall HOTL control problem can be summarized as follows:

**Objective:** Design a control law $u_{c_i}(t)$ such that under normal operating conditions, system states remain bounded and asymptotically converge, while during operating contingencies the system preserves sufficient GRC with minimal CPD.

**Challenges:** The main challenges include $(i)$ accounting for inverter power capacity limits, $(ii)$ adapting to uncertain or partial contingency information, and $(iii)$ ensuring coordinated decision-making between automated control and human operator.

To address these challenges, the proposed HOTL architecture integrates:

1. An **adaptive controller layer**, embedded in the DERMS, that performs fast, real-time control of IBRs' active and reactive power injections while accounting for saturation and model uncertainties; and
2. A **human operator,** who performs supervisory cognitive tasks—such as degradation identification, degradation estimation $\hat{\Lambda}_f$, and tuning of the resilience trade-off parameter $\mu$—to guide the adaptive controller toward resilient operation.

In this framework, the adaptive control provides high-speed response capability, while the operator offers contextual awareness and judgment, forming a human-on-the-loop resilient control system. Detailed architecture and control design are presented in Section III.

## III. HOTL Control Architecture

Building upon the formulation in Section II, this section details the HOTL control structure that enables resilient operation of IBRs. The proposed architecture incorporates a two-layer hierarchy in which an adaptive control layer ensures fast, autonomous adjustment of IBR active and reactive power injections, while a human-supervisory layer adjusts control objectives in response to contingencies. The human operator, through DERMS interface, provides supervisory input including degradation estimation (and the resilience tuning parameter ($\mu$).

The interaction between these two layers establishes a closed loop supervisory control framework that maintains real-time voltage regulation while explicitly managing the trade-off between GRC and CPD. The following subsections describe the adaptive control law and its supervisory coordination mechanisms.

### A. *μ–Mod Adaptive Controller for Inverter Network*

The adaptive controller is designed to ensure stable tracking performance of IBRs despite system nonlinearities, inverter power capacity constraints, and parametric uncertainties.

**Reference Model:** The desired dynamic behavior of the voltage profile in an active distribution network is represented by the reference model [30], [31]:

$$\begin{aligned} \dot{x}_m(t) &= A_m x_m(t) + B_m r_0(t) \\ y_m(t) &= C x_m(t) \end{aligned} \tag{3}$$

where $r_0(t)$ is the desired reference, and $A_m$ is a Hurwitz matrix. $x_m \in R^n$ is the state of the reference model, $(A_m, B_m)$ is controllable, and $C$ is defined as given in (1), that is, $y_m \in R^k$ corresponds to a reference model output. The reference model represents the nominal operating trajectory of the network under normal conditions.

***μ*–Mod Adaptive Control Law:** The objective of the adaptive controller is to determine the control input $u_c$ so that the state deviation error $e(t) = x(t) - x_m(t)$ converges to zero asymptotically and all the signals in the adaptive system remain bounded. The proposed adaptive control is a combined $\mu$-mod and closed-loop reference model (CRM) adaptive control as expressed in (4):

$$u_c(t) = \begin{cases} u_{ad}(t), & |u_{ad}(t)| \le u_{max}^{\delta} \\ \dfrac{1}{1+\mu}\left(u_{ad}(t) + \mu\, sgn(u_{ad}(t)) u_{max}^{\delta}\right) & |u_{ad_i}(t)| \ge u_{max}^{\delta} \end{cases} \tag{4}$$

where

$$u_{ad_i}(t) = K_x^T(t)x(t) + K_r^T(t)r_0(t) + \hat{d}(t) + \hat{\Phi}^{\mathsf{T}}(t) f(x) \tag{5}$$

and the virtual saturation limit is defined as

$$u_{max_i}^{\delta} = (1-\delta) u_{max_i}, \quad 0 \le \delta < 1 \tag{6}$$

where $\delta$ introduces a buffer region $[(1-\delta)u_{max_i}, u_{max_i}]$ between the nominal and physical limits, ensuring that the inverter operates safely within its capacity bounds. The parameter $\mu$ determines how aggressively the control effort approaches this limit: larger $\mu$ values correspond to more conservative operation, preserving greater GRC but allowing larger CPD.

**Modified Reference Model under Contingency**: Following a contingency, the reference model is revised to account for the degraded system capacity resulting from reduced available generation capacity and inverter power limitations:

$$\dot{x}_m(t) = A_m x_m(t) + B_m\left(r_0(t) + K_u^T(t)\Delta u_{ad}(t)\right) - Le(t) \tag{7}$$

$$\Delta u_{ad}(t) = u_{max}\, sat(u_c(t)/u_{max}) - u_{ad}(t) \tag{8}$$

where $e(t) = x(t) - x_m(t)$ denotes the state deviation error and $L < 0$ is chosen such that $A_m + L$ is Hurwitz. $K_u^T(t)$ is the feedforward compensation for reference tracking. This structure allows the reference model to adapt dynamically in response to inverter power capacity constraints, thereby enabling *graceful degradation* instead of instability under constrained operation conditions.

**Adaptive Parameter Update Laws:** The adaptive gains are updated using the following laws:

$$\begin{aligned} \dot{K}_x(t) &= -\Gamma_x x(t) e^T(t) P B \\ \dot{K}_r(t) &= -\Gamma_r r_0(t) e^T(t) P B \\ \dot{K}_u(t) &= \Gamma_u \Delta u_{ad} e^T(t) P B_m \\ \dot{\hat{d}}(t) &= -\Gamma_d e^T(t) P B \\ \dot{\hat{\Phi}}(t) &= -\Gamma_f f(x(t)) e^T(t) P B \end{aligned} \tag{9}$$

where $P = P^T > 0$ satisfies the Lyapunov equation $A_m^T P + P A_m = -Q$ for $Q > 0$, and the adaptation gains $\Gamma_x, \Gamma_r, \Gamma_u, \Gamma_d, \Gamma_f > 0$. The adaptive inverter controller described by (1)–(6) provides the control input $u_c(t)$ in (4) as a real-time solution to the resilient regulation problem defined in Section II. The controller consists of a combination of fixed and tunable parameters that determine the controller's dynamic and resilience characteristics. These include the resilience parameters $\mu$ and $\delta$, the reference model parameters $A_m, B_m, L$, and the adaptive controller initial conditions $K_x(0), K_r(0), K_u(0), \hat{d}(0), \widehat{\Phi}(0)$. The choice of the buffer coefficient $\delta$ sets the desired GRC margin, as defined in (6). A smaller $\delta$ reduces the buffer region and allows the inverter to operate closer to its physical limits, thereby reducing available reserve capacity. Conversely, a larger $\delta$ increases GRC but may restrict dynamic response. The resilience tuning parameter $\mu$ has a proportional influence on both GRC and CPD; increasing $\mu$ enhances reserve preservation but permits higher tracking deviation. In addition to $\mu$ and $\delta$, the adaptive controller requires the reference model parameters $A_m, B_m, L$ and the initial control parameters $K_x(0), K_r(0), K_u(0)$. When no contingencies are present ($\Lambda_f = I$), the matrices $A_m$ and $B_m$, as well as the control parameters, are initialized as (10):

$$\begin{gathered} A_m = A + BK_x^T(0) \\ K_r^T(0) = -(A_m^{-1}B)^{-1} \\ K_u^T(0) = -A_m^{-1}B \\ B_m = BK_r^T(0) \end{gathered} \tag{10}$$

Here, $K_x(0)$ is obtained using a linear–quadratic regulator (LQR) approach based on the nominal plant matrices $(A, B)$, ensuring an optimal trade-off between state tracking and control effort.

**Quantitative Resilience Metrics:** The resilience metrics, GRC and CPD, are quantitatively defined according to (11)-(12) to assess and tune the system's resilience:

$$GRC = \max_i x(u_{max_i} - u^{\delta}_{max_i})/rms\,\left(min\left(u_{max_i} - |u_i\,(t)|\right)\right)\Big|_{t_a}^{T} \tag{11}$$

$$CPD = rms(y_m - r_0(t))/rms(r_0(t))\,,\, t \in T_0 \tag{12}$$

where $t_a$ and $T$ denote the start and end of the degradation period, and $T_0$ is the observation window for post-contingency performance. As $\mu$ increases, GRC improves, but CPD also increases due to more conservative control action. Therefore, $\mu$ represents the resilience tuning parameter, enabling real-time adjustment of the balance between reserve capacity and performance degradation.

*B. Human Operator and Supervisory Role*

The human operator forms the upper layer of the proposed control hierarchy and contributes to the system's resilience through supervisory assessment, decision-making, and parameter tuning. While the adaptive controller executes rapid, automated regulation, the operator maintains strategic oversight by analyzing situational data and adjusting high-level parameters that govern system behavior during contingencies. When a contingency or abnormal event is detected, the operator performs two primary tasks:

**Degradation Assessment and Estimation**: Using situational awareness, operational experience, contextual judgment, and available system information, the operator assesses the severity and location of the contingency and contributes to the estimation of the degradation matrix $\hat{\Lambda}_f$. This matrix quantifies the reduction in power capacity of individual inverter or IBR unit resulting from equipment faults, cyber-physical attacks, weather conditions, or other adverse operating conditions. Based on the operator's evaluation, the adaptive controller updates its internal reference model parameters as (13):

$$\begin{gathered} A_m = A + B\hat{\Lambda}_f K_x^T(t_a) \\ K_r^T(t_a) = -\left(A_m^{-1}B\hat{\Lambda}_f(t_a)\right)^{-1} \\ B_m = B\hat{\Lambda}_f K_r^T(t_a) \\ K_u^T(t_a) = -A_m^{-1}B\hat{\Lambda}_f(t_a) \end{gathered} \tag{13}$$

where $t_a$ denotes the degradation onset time. These updates enable the adaptive controller to incorporate the operator's situational awareness and maintain model consistency under degraded operating conditions. To represent the variability and reliability of the human operator's degradation assessment, a confidence factor $\eta(t) \in (0,1]$ is introduced. This coefficient represents an evolving trust state that the adaptive controller places in the operator's degradation assessment, formed and updated over time based on the observed outcome of relying on that assessment. A generalized expression for $\eta(t)$ can be defined as (14):

$$\eta(t) = f(OE) \tag{14}$$

where OE denotes the operator's experience index, such as the number of previously managed contingency events, accumulated operational hours, or historical system reliability performance metrics. The normalization function $f(.)$ maps these experience indicators into the range (0,1], ensuring consistent scaling across different operators or system configurations while allowing η(t) to increase when the operator's judgment has led to favorable outcomes and decrease when it has not. The final evaluation of the degradation matrix used by the adaptive controller is then expressed as

$$\hat{\Lambda}_f = \eta(t)\,\hat{\Lambda}_{f,o} + (1-\eta)\Lambda_{nom} \tag{15}$$

where $\hat{\Lambda}_{f,o}$ represent the operator's evaluation of the loss of control effectiveness derived from real-time data analytics and system-level diagnostic tools. $\Lambda_{nom} = I$ denotes the nominal condition. When $\eta(t)$ approaches 1, the controller relies primarily on the operator's degradation assessment, indicating high confidence or strong situational awareness. Conversely, smaller values of $\eta(t)$ reduce the influence of the operator's evaluation, causing the controller to rely more heavily on nominal system assumptions.

**Specification of the Resilience Parameter $\boldsymbol{\mu}$**: The operator determines the resilience tuning parameter $\mu$, which governs the trade-off between GRC and CPD. A larger $\mu$ increases GRC by maintaining higher reserve margins but leads to greater CPD (i.e., looser voltage tracking), whereas a smaller $\mu$ prioritizes tracking accuracy over the preservation of reserve capacity. The selection of $\mu$ thus reflects the operator's judgment, informed by experience, contingency situation, and operational objectives.

After these supervisory inputs ($\mu$ and $\widehat{\Lambda}_f$) are specified, the adaptive controller autonomously executes control actions according to (1)–(15). This human–automation teaming ensures that control authority is appropriately shared: the automation manages real-time inverter dynamics, while the operator supervises system operating conditions, evaluates contingency severity, and adjusts the resilience parameter to preserve inverter reserve margin and system stability.

*C. Stability Analysis*

Theorem 1 establishes the stability of the proposed controller.

**Theorem 1 (Stability of the Proposed HOTL µ-Mod Adaptive Controller):** Consider the closed-loop system described by (1)–(15). Under the proposed HOTL µ-mod adaptive controller, all signals in the adaptive system remain bounded, the tracking error $e(t) = x(t) - x_{\mathrm{m}}(t)$ converges to zero asymptotically, and the control input satisfies $|u_{ci}| \leq U^{\delta}_{max_i}$ for all t > 0, preventing actuator saturation.

**Proof:** Define the tracking error as:

$$e(t) = x(t) - x_m\ (t) \tag{16}$$

Substituting the plant dynamics (1) and the reference model (9) into the error definition, the tracking error dynamics are obtained as:

$$\begin{aligned} \dot{e}(t) = Ax(t) + B\Lambda_f u\,(t) + d + \Phi^T f(x) - A_m x_m\,(t) \\ -B_m[r_0(t) + K_u^T(t)\Delta u_{ad}(t)] + Le(t). \end{aligned} \tag{17}$$

Following the standard matching condition used in model reference adaptive control, let $K_x^*$, $K_r^*$, $K_u^*$, $d^*$, and $\Phi^*$ denote the ideal parameters satisfying:

$$\begin{aligned} A + B\Lambda_f (K_x^*)^T = A_m \\ B\Lambda_f (K_r^*)^T = B_m \\ B\Lambda_f d^* = d \\ B\Lambda_f (\Phi^*)^T = \Phi^T \\ B_m (K_u^*)^T = B\Lambda_f \end{aligned} \tag{18}$$

Define the parameter estimation errors as

$$\begin{aligned} \widetilde{K}_x(t) = K_x(t) - K_x^* \\ \widetilde{K}_r(t) = K_r(t) - K_r^* \\ \widetilde{K}_u(t) = K_u(t) - K_u^* \\ \tilde{d}(t) = \hat{d}(t) - d^* \\ \widetilde{\Phi}(t) = \widehat{\Phi}(t) - \Phi^* \end{aligned} \tag{19}$$

From (2) and the definition of $\Delta u_{ad}(t)$ in (10), it follows that $u\,(t) = u_{ad}(t) + \Delta u_{ad}(t)$. Using the adaptive control law (5), the term $B\Lambda_f u\,(t)$ in (17) can be expanded as:

$$\begin{aligned} B\Lambda_f u\,(t) = B\Lambda_f K_x^T(t)x(t) + B\Lambda_f K_r^T(t)r_0(t) \\ + B\Lambda_f \hat{d}(t) + B\Lambda_f \widehat{\Phi}^T(t)f(x) \\ + B\Lambda_f \Delta u_{ad}(t) \end{aligned} \tag{20}$$

Substituting the parameter estimation error definitions from (19) into (20) and enforcing the matching conditions in (18), yields

$$\begin{aligned} B\Lambda_f u\,(t) = (A_m - A)x(t) + B\Lambda_f \widetilde{K}_x^T(t)x(t) \\ +B_m r_0(t) + B\Lambda_f \widetilde{K}_r^T(t)r_0(t) + d + B\Lambda_f \tilde{d}(t) \\ +\Phi^T f(x) + B\Lambda_f \widetilde{\Phi}^T(t)f(x) + B\Lambda_f \Delta u_{ad}(t) \end{aligned} \tag{21}$$

Given that $K_u(t) = K_u^* + \widetilde{K}_u(t)$, the term $-B_m K_u^T(t)\Delta u_{ad}(t)$ in (17) can be expanded as:

$$\begin{aligned} -B_m K_u^T(t)\Delta u_{ad}(t) &= -B_m (K_u^*)^T \Delta u_{ad}(t) \\ &\quad -B_m \widetilde{K}_u^T(t)\Delta u_{ad}(t). \end{aligned} \tag{22}$$

Substituting (21) and (22) into (17), and using the matching condition $B_m(K_u^*)^T = B\Lambda_f$, gives

$$\begin{aligned} \dot{e}(t) = (A_m + L)e(t) + B\Lambda_f \widetilde{K}_x^T(t)x(t) \\ +B\Lambda_f \widetilde{K}_r^T(t)r_0(t) + B\Lambda_f \tilde{d}(t) + B\Lambda_f \widetilde{\Phi}^T(t)f(x) \\ -B_m \widetilde{K}_u^T(t)\Delta u_{ad}(t). \end{aligned} \tag{23}$$

Defining $A_L = A_m + L$, the tracking error dynamics can be written compactly as

$$\begin{aligned} \dot{e}(t) = A_L e(t) + B\Lambda_f \widetilde{K}_x^T(t)x(t) + B\Lambda_f \widetilde{K}_r^T(t)r_0(t) \\ +B\Lambda_f \tilde{d}(t) + B\Lambda_f \widetilde{\Phi}^T(t)f(x) - B_m \widetilde{K}_u^T(t)\Delta u_{ad}(t). \end{aligned} \tag{24}$$

Remark 1: The structure of (24) forms the foundation for the subsequent stability analysis. First, the CRM gain $L$ is selected such that $A_L = A_m + L$ is Hurwitz. Consequently, the system $\dot{e}(t) = A_L e(t)$ is exponentially stable. Second, all uncertainties appear only through the parameter estimation errors $\widetilde{K}_x$, $\widetilde{K}_r$, $\widetilde{K}_u$, $\tilde{d}$, and $\widetilde{\Phi}$. This structure enables cancellation of the cross terms in the Lyapunov derivative. Finally, the degradation matrix $\Lambda_f$ appears explicitly in the error dynamics, ensuring that the adaptive controller accounts for reduced inverter capacity under contingency conditions.

Since $A_L = A_m + L$ is Hurwitz, for any symmetric positive definite matrix $Q = Q^T > 0$, there exists a unique positive definite matrix $P = P^T > 0$ satisfying.

$$A_L^T P + PA_L = -Q. \tag{25}$$

Consider the Lyapunov function candidate

$$\begin{aligned} V = e^T Pe + tr(\Lambda_f \widetilde{K}_x^T \Gamma_x^{-1} \widetilde{K}_x) + tr\ (\Lambda_f \widetilde{K}_r^T \Gamma_r^{-1} \widetilde{K}_r) \\ +\Lambda_f \tilde{d}^T \Gamma_d^{-1} \tilde{d} + tr\ (\Lambda_f \widetilde{\Phi}^T \Gamma_f^{-1} \widetilde{\Phi}) + tr\ (\widetilde{K}_u^T \Gamma_u^{-1} \widetilde{K}_u) \end{aligned} \tag{26}$$

Taking the time derivative of $V$ along the trajectories of (24) and substituting the adaptive update laws yields cancellation of all parameter-error cross terms. Therefore,

$$\dot{V} = e^T(A_L^T P + PA_L)e \tag{27}$$

Using (25), it follows that

$$\dot{V} = -e^T Qe \leq 0 \tag{28}$$

Hence, $V(t)$ is non-increasing and satisfies

$$V(t) \leq V(0) < \infty, \forall t \geq 0 \tag{29}$$

Therefore,

$$e(t), \widetilde{K}_x(t), \widetilde{K}_r(t), \widetilde{K}_u(t), \tilde{d}(t), \widetilde{\Phi}(t) \in L_\infty \tag{30}$$

Integrating (28) over $[0, t]$ gives

$$\int_0^t e^T\,(\tau)Qe(\tau)d\tau = V(0) - V(t) \leq V(0). \tag{31}$$

Since $Q > 0$,

$$\int_0^\infty \parallel e(\tau) \parallel^2 d\tau < \infty \tag{32}$$

which implies

$$e(t) \in L_2 \tag{33}$$

Consequently, from (30) and (33):

$$e(t) \in L_2 \cap L_\infty \tag{34}$$

Remark 2: Here, $L_\infty$ denotes the set of bounded signals, while $L_2$ denotes the set of finite-energy signals. Therefore, (30) establishes boundedness of the tracking error and adaptive parameter estimation errors, whereas (33) shows that the tracking error possesses finite energy. These properties are used in the subsequent convergence analysis.

Since $e(t) \in L_2 \cap L_\infty$, and since all adaptive parameters and system signals appearing in (24) are bounded, it follows that

$$\dot{e}(t) \in L_\infty. \tag{35}$$

Therefore, $e(t)$ is uniformly continuous.

Lemma 1 (Barbalat's Lemma [32]): Suppose $f: [0, \infty) \rightarrow \mathbb{R}$ is uniformly continuous and $\lim_{t\to\infty} \int_0^t f(\tau)\, d\tau$ exists and is finite. Then $\lim_{t\to\infty} f(t) = 0$. The proof of this lemma is omitted for brevity and can be found in [32].

By Barbalat's Lemma, since $e(t) \in L_2$ and $\dot{e}(t) \in L_\infty$, the tracking error satisfies

$$\lim_{t\to\infty} e(t) = 0. \tag{36}$$

Thus,

$$\lim_{t\to\infty} (x(t) - x_m(t)) = 0. \tag{36}$$

Therefore, the plant state asymptotically tracks the modified reference model state.

Since $r_0(t)$, $K_u(t)$, and $\Delta u_{ad}(t)$ are bounded, the input to the modified reference model is bounded. Therefore, using (9) and the Hurwitz property of $A_m$, it follows that $x_m(t) \in L_\infty$. Given $x(t) = e(t) + x_m(t)$ and $e(t) \in L_\infty$, it can be concluded that $x(t) \in L_\infty$. Since $x(t)$, $r_0(t)$, $\hat{d}(t)$, $\hat{\Phi}(t)$, and all adaptive parameters are bounded, the adaptive control signal $u_{ad}(t)$ is also bounded. Furthermore, the μ-mod saturation law guarantees that the applied control signal remains bounded. That is,

$$u_c(t) \in L_\infty \tag{37}$$

To prove the saturation-prevention property, consider the $i$-th control channel. If $| u_{ad}(t) | \leq u_{max}^{\delta}$, then from (4), $u_c(t) = u_{ad}(t)$, and therefore $| u_c(t) | \leq u_{max,}^{\delta} < u_{max,}$.

If $| u_{ad}(t) | > u_{max,}^{\delta}$,

$$u_c(t) = \frac{1}{1+\mu}\left(u_{ad}(t) + \mu\, \mathrm{sgn}(u_{ad}(t)) u_{max}^{\delta}\right) \tag{37}$$

Taking the absolute value of (37) gives:

$$| u_c(t) | \leq \frac{1}{1+\mu}\left[| u_{ad}(t) | + \mu u_{max}^{\delta}\right]. \tag{38}$$

Since $u_{ad}(t) \in L_\infty$, there exists a finite constant $\bar{u}_{ad,i}$ such that $| u_{ad}(t) | \leq \bar{u}_{ad}, \forall t \geq 0$. Thus, $| u_c(t) | \leq \frac{\bar{u}_{ad}+\mu u_{max}^{\delta}}{1+\mu}$. Therefore, to ensure that the physical actuator limits are not violated: $\mu$ can be chosen according to (39):

$$\frac{\bar{u}_{ad} + \mu u_{max}^{\delta}}{1+\mu} \leq u_{max}. \tag{39}$$

Solving (39) gives

$$\mu \geq \frac{\bar{u}_{ad} - u_{max}}{u_{max} - u_{max}^{\delta}} \tag{40}$$

Therefore, if the human operator adjusts μ such that

$$\mu \geq max\left\{0, \frac{\bar{u}_{ad} - u_{max}}{u_{max} - u_{max}^{\delta}}\right\}. \tag{41}$$

Then, one can have:

$$| u_c(t) | \leq u_{max}, \forall t \geq 0. \tag{42}$$

Thus, actuator saturation is prevented. This completes the proof.

**Remark 3.** The bound $\bar{u}_{ad,i}$ used in (40)–(41) depends on the accuracy of the operator's degradation estimate. Recall that, at the degradation onset $t_a$, the reference model $A_m$ in (13) is constructed using the operator-supplied estimate $\widehat{\Lambda}_f$, whereas the ideal parameters $K_x^*$, $K_r^*$, $K_u^*$, $d^*$, and $\Phi^*$ satisfy the matching conditions in (18) with respect to the true degradation matrix $\Lambda_f$. Define the degradation-estimation gap as

$$\Delta\Lambda_f \triangleq \Lambda_f - \widehat{\Lambda}_f \tag{43}$$

Accordingly, the initial parameter errors at $t_a$ are given by

$$\begin{aligned} \widetilde{K}_x(t_a) &= K_x(t_a)\Lambda_f^{-1}\Delta\Lambda_f, \\ \widetilde{K}_r(t_a) &= K_r(t_a)\Lambda_f^{-1}\Delta\Lambda_f, \\ \widetilde{K}_u(t_a) &= K_u(t_a)\Lambda_f^{-1}\Delta\Lambda_f, \\ \tilde{d}(t_a) &= d(t_a)\Lambda_f^{-1}\Delta\Lambda_f, \\ \widetilde{\Phi}(t_a) &= \Phi(t_a)\Lambda_f^{-1}\Delta\Lambda_f. \end{aligned} \tag{44}$$

Since $\dot{V} \leq 0$ for all $t \geq t_a$, according to (28), the Lyapunov function satisfies

$$V(t) \leq V(t_a),\ t \geq t_a \tag{45}$$

Moreover, every term in the Lyapunov function (26) is nonnegative. Therefore,

$$\mathrm{tr}\left(\Lambda_f \widetilde{K}_x^T(t)\Gamma_x^{-1}\widetilde{K}_x(t)\right) \leq V(t_a) \tag{46}$$

which implies

$$\|\widetilde{K}_x(t)\| \leq \sqrt{\frac{\lambda_{\max}(\Gamma_x)}{\lambda_{\min}(\Lambda_f)} V(t_a)} \triangleq k_x^{\max} \tag{47}$$

Similarly,

$$\begin{aligned} \|\widetilde{K}_r(t)\| &\leq \sqrt{\frac{\lambda_{\max}(\Gamma_r)}{\lambda_{\min}(\Lambda_f)} V(t_a)} \triangleq k_r^{\max}, \\ \|\widetilde{K}_u(t)\| &\leq \sqrt{\frac{\lambda_{\max}(\Gamma_u)}{\lambda_{\min}(\Lambda_f)} V(t_a)} \triangleq k_u^{\max}, \\ \|\tilde{d}(t)\| &\leq \sqrt{\frac{\lambda_{\max}(\Gamma_d)}{\lambda_{\min}(\Lambda_f)} V(t_a)} \triangleq k_d^{\max}, \\ \|\widetilde{\Phi}(t)\| &\leq \sqrt{\frac{\lambda_{\max}(\Gamma_\Phi)}{\lambda_{\min}(\Lambda_f)} V(t_a)} \triangleq k_\Phi^{\max}. \end{aligned} \tag{48}$$

Substituting (48) into the Lyapunov function (26), and using

$$\|M\Lambda_f^{-1}\Delta\Lambda_f\| \leq \frac{\| M \|_F}{\lambda_{\min}(\Lambda_f)}\|\Delta\Lambda_f\| \tag{49}$$

gives the complete bound

$$V(t_a) \le \frac{\lambda_{\max}(\Lambda_f)}{\lambda_{\min}(\Gamma_x)\lambda_{\min}^2(\Lambda_f)}\|K_x(t_a)\|^2\|\Delta\Lambda_f\|^2 + \frac{\lambda_{\max}(\Lambda_f)}{\lambda_{\min}(\Gamma_r)\lambda_{\min}^2(\Lambda_f)}\|K_r(t_a)\|^2\|\Delta\Lambda_f\|^2 + \frac{\lambda_{\max}(\Lambda_f)}{\lambda_{\min}(\Gamma_u)\lambda_{\min}^2(\Lambda_f)}\|K_u(t_a)\|^2\|\Delta\Lambda_f\|^2 + \frac{\lambda_{\max}(\Lambda_f)}{\lambda_{\min}(\Gamma_d)\lambda_{\min}^2(\Lambda_f)}\|d(t_a)\|^2\|\Delta\Lambda_f\|^2 + \frac{\lambda_{\max}(\Lambda_f)}{\lambda_{\min}(\Gamma_\Phi)\lambda_{\min}^2(\Lambda_f)}\|\Phi(t_a)\|^2\|\Delta\Lambda_f\|^2. \quad (50)$$

Equivalently, define

$$C_V \triangleq \frac{\lambda_{\max}(\Lambda_f)}{\lambda_{\min}^2(\Lambda_f)}\left[\frac{\| K_x(t_a) \|^2}{\lambda_{\min}(\Gamma_x)} + \frac{\| K_r(t_a) \|^2}{\lambda_{\min}(\Gamma_r)} + \frac{\| K_u(t_a) \|^2}{\lambda_{\min}(\Gamma_u)} + \frac{\| d(t_a) \|^2}{\lambda_{\min}(\Gamma_d)} + \frac{\| \Phi(t_a) \|^2}{\lambda_{\min}(\Gamma_\Phi)}\right]. \quad (51)$$

Then,

$$(t_a) \le C_V\|\Delta\Lambda_f\|^2 \quad (51)$$

Consequently, the parameter-error bounds can be expressed as

$$k_x^{\max} \le c_x\|\Delta\Lambda_f\|, \quad c_x \triangleq \sqrt{\frac{\lambda_{\max}(\Gamma_x)}{\lambda_{\min}(\Lambda_f)}C_V},$$
$$k_r^{\max} \le c_r\|\Delta\Lambda_f\|, \quad c_r \triangleq \sqrt{\frac{\lambda_{\max}(\Gamma_r)}{\lambda_{\min}(\Lambda_f)}C_V},$$
$$k_u^{\max} \le c_u\|\Delta\Lambda_f\|, \quad c_u \triangleq \sqrt{\frac{\lambda_{\max}(\Gamma_u)}{\lambda_{\min}(\Lambda_f)}C_V}, \quad (52)$$
$$k_d^{\max} \le c_d\|\Delta\Lambda_f\|, \quad c_d \triangleq \sqrt{\frac{\lambda_{\max}(\Gamma_d)}{\lambda_{\min}(\Lambda_f)}C_V},$$
$$k_\Phi^{\max} \le c_\Phi\|\Delta\Lambda_f\|, \quad c_\Phi \triangleq \sqrt{\frac{\lambda_{\max}(\Gamma_\Phi)}{\lambda_{\min}(\Lambda_f)}C_V}.$$

Using the adaptive control law (5) together with these bounds, the adaptive control input satisfies

$$\bar{u}_{ad} = (\|K_x^*\| + k_x^{\max}) \| x \|_\infty + (\|K_r^*\| + k_r^{\max})r_{0,\max} + (|d^*| + k_d^{\max}) + (\|\Phi^*\| + k_\Phi^{\max}) \| f(x) \|_\infty + (\|K_u^*\| + k_u^{\max}) \| \Delta u_c \|_\infty. \quad (53)$$

Substituting (52) into (53) yields

$$\bar{u}_{ad} \le \left(\|K_x^*\| + c_x \| \Delta\Lambda_f \|\right) \| x \|_\infty + \left(\|K_r^*\| + c_r \| \Delta\Lambda_f \|\right)r_{0,\max} + \left(|d^*| + c_d \| \Delta\Lambda_f \|\right) + \left(\|\Phi^*\| + c_\Phi \| \Delta\Lambda_f \|\right) \| f(x) \|_\infty + \left(\|K_u^*\| + c_u \| \Delta\Lambda_f \|\right) \| \Delta u_c \|_\infty. \quad (54)$$

Therefore, substituting this result into (41), the minimum admissible value of $\mu$ satisfies

$$\mu \ge \max\{0, \frac{1}{u_{\max,} - u_{\max,}^{\delta}}[\left(\|K_x^*\| + c_x \| \Delta\Lambda_f \|\right) \| x \|_\infty + \left(\|K_r^*\| + c_r \| \Delta\Lambda_f \|_F\right)r_{0,\max} + \left(|d^*| + c_d \| \Delta\Lambda_f \|\right) + \left(\|\Phi^*\| + c_\Phi \| \Delta\Lambda_f \|\right) \| f(x) \|_\infty + \left(\|K_u^*\| + c_u \| \Delta\Lambda_f \|\right) \| \Delta u_c \|_\infty - u_{\max,}]\}. \quad (55)$$

Every term depending on $\Delta\Lambda_f$ in (55) has a nonnegative coefficient. Therefore, a larger discrepancy between the operator-supplied estimate $\widehat{\Lambda}_f$ and the true actuator degradation $\Lambda_f$ increases the minimum admissible value of $\mu$. This establishes a direct coupling between human-supervisory estimation quality and the resilience–tracking tradeoff. A less accurate degradation estimate requires a larger value of $\mu$, which, results in a less reserve bound.

## IV. Comparison With Other Control Strategies

In order to evaluate the effectiveness of the proposed HOTL control framework, this section reviews a conventional fault-tolerant control (FTC) method.

### A. Fault tolerant control

In previous studies such as [33], a fault-tolerant control (FTC) strategy was developed to mitigate actuator-related faults and performance degradation in control systems. The FTC framework is composed of two primary components: a fault-diagnosis module and a control reconfiguration module. The diagnosis module assumes that the impact of a contingence can be identified through estimation techniques such as banks of Kalman filters or parameter observers. Once the abnormal situation is detected and its magnitude estimated, the control module adjusts its feedback structure accordingly. This controller typically employs fixed feedback and feedforward gains along with an updated reference model, which together ensures stable operation after fault isolation. When degradation in actuator performance is detected, the corresponding control signals are intentionally reduced in magnitude to prevent saturation and maintain bounded responses. A detailed description of this FTC methodology and its stability analysis can be found in [33].

## V. Simulation Studies and Results

The proposed human-on-the-loop adaptive control architecture was evaluated on a grid-connected inverter subject to sequential actuator degradation. The test system, illustrated in Fig. 2, consists of a grid-connected inverter integrated to the grid at the point of common coupling (PCC). The system consists of two inputs and one output, where the plant state is defined as $x = [V]$, representing the voltage magnitude at the PCC bus. The control input $u = [P, Q]$ corresponds to active and reactive power of inverter. The controller adjusts both active $(P)$ and reactive $(Q)$ power to achieve stable tracking of the voltage $(V\angle\theta)$, while respecting physical inverter power capacity limits and preserving GRC under degraded operating conditions.

To evaluate controller performance under contingency conditions, the inverter is assumed to undergo two sequential degradation events. During the first event, occurring at t = 100 s, the power capacity of the inverter actuators is reduced by 50%. Then, the second degradation of 25% is added at t = 150 s, resulting in a cumulative loss in inverter capacity. These degradation events are represented by the effectiveness matrix $\Lambda_f$. In addition to actuator degradation, a load disturbance is introduced at $t = 200$s, representing a sudden increase in power demand at the PCC. two control frameworks are compared:

1. Human-on-the-loop adaptive controller
2. Fault tolerant controller

The control input is constrained by a physical saturation limit of $U_{max} = 0.7\,pu$, with a virtual saturation threshold defined by (16):

$$U_\delta = (1-\delta)U_{\max}, \qquad \delta = 0.15 \tag{56}$$

resulting in a virtual constraint of $U_{max}^{\delta} = 0.6\,pu$, preserving a 15% inverter reserve capacity, thereby maintaining sufficient reserve capacity to support resilient operation under contingencies.

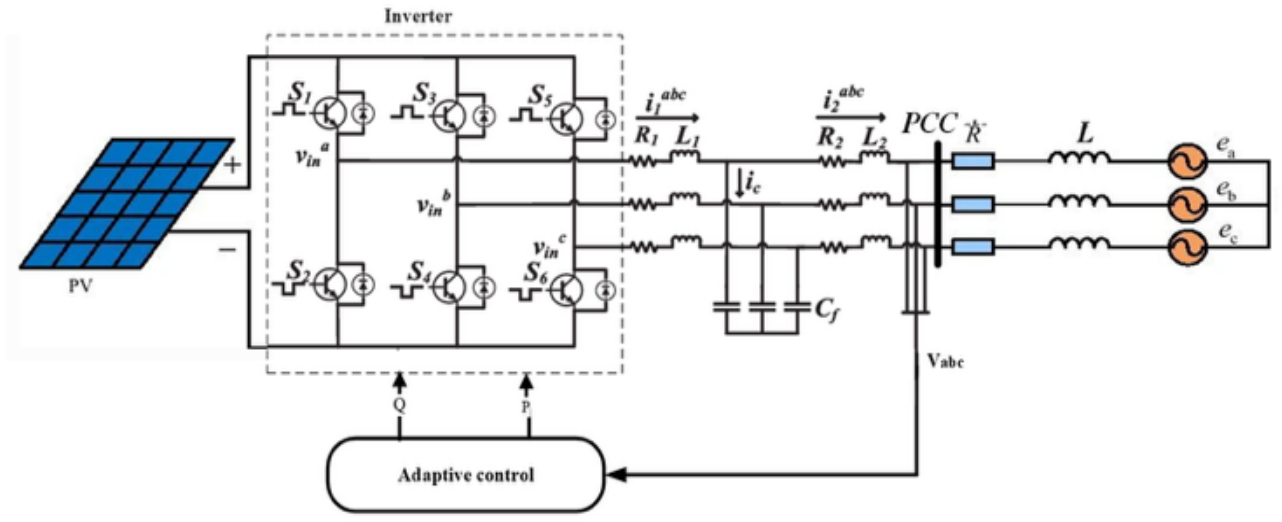

**Fig. 2.** A grid-connected inverter structure.

*A. Human-on-the-loop Adaptive Controller Response*

Under the human-on-the-loop control framework, the operator performs two tasks: anomaly assessment and tuning of the resilience parameter $\mu$. At t = 100 s, after identifying the 50% degradation, the operator provides an estimate of the effectiveness matrix $\hat{\Lambda}_f$ and selects $\mu = 1.2$ based on prior operational experience with similar degradation levels. This selection enables the controller to prioritize tracking performance while adjusting the reference model to address reduced actuator capacity.

Fig. 3 presents the results obtained from applying the proposed HOTL controller. Fig. 3 (a) displays the voltage magnitude, denoted ($V_2$). Fig 3 (b) and (c) illustrate the control inputs, with ($P_2$) and ($Q_2$) indicating active and reactive power of the inverter.

As shown in Fig. 3 (a)–(c), the $V_2$ trajectories is well regulated with negligible overshoot, and the control inputs ($P_2, Q_2$) maintain a safe distance from the saturation boundary. At the second event, t = 150 s, the operator observes increased stress on the inverter and increases the resilience parameter to $\mu = 10$ according to his experience of new situation. This change makes the controller substantially more conservative—increasing active and reactive power injection ($P_2, Q_2$) while preserving inverter reserve capacity and preventing operation near the physical saturation limits. This intervention is essential to maintain controllability of the IBR following severe degradation. The increased $\mu$ also induces controlled performance degradation, reflected in reduced voltage-tracking aggressiveness, however, still maintains $V_2$ within allowable voltage limits. At $t = 200$s, a load change is introduced, resulting in an increase in both active and reactive power demand. Despite this additional disturbance, the proposed controller by utilizing the inverter reserve capacity is able to compensate for the increased load. As a result, the voltage $V_2$ remains within the acceptable operating range.

These results confirm that human supervisory input enables dynamic adjustment of the control strategy, improving the system's ability for resilience operation and robust voltage regulation under actuator degradation.

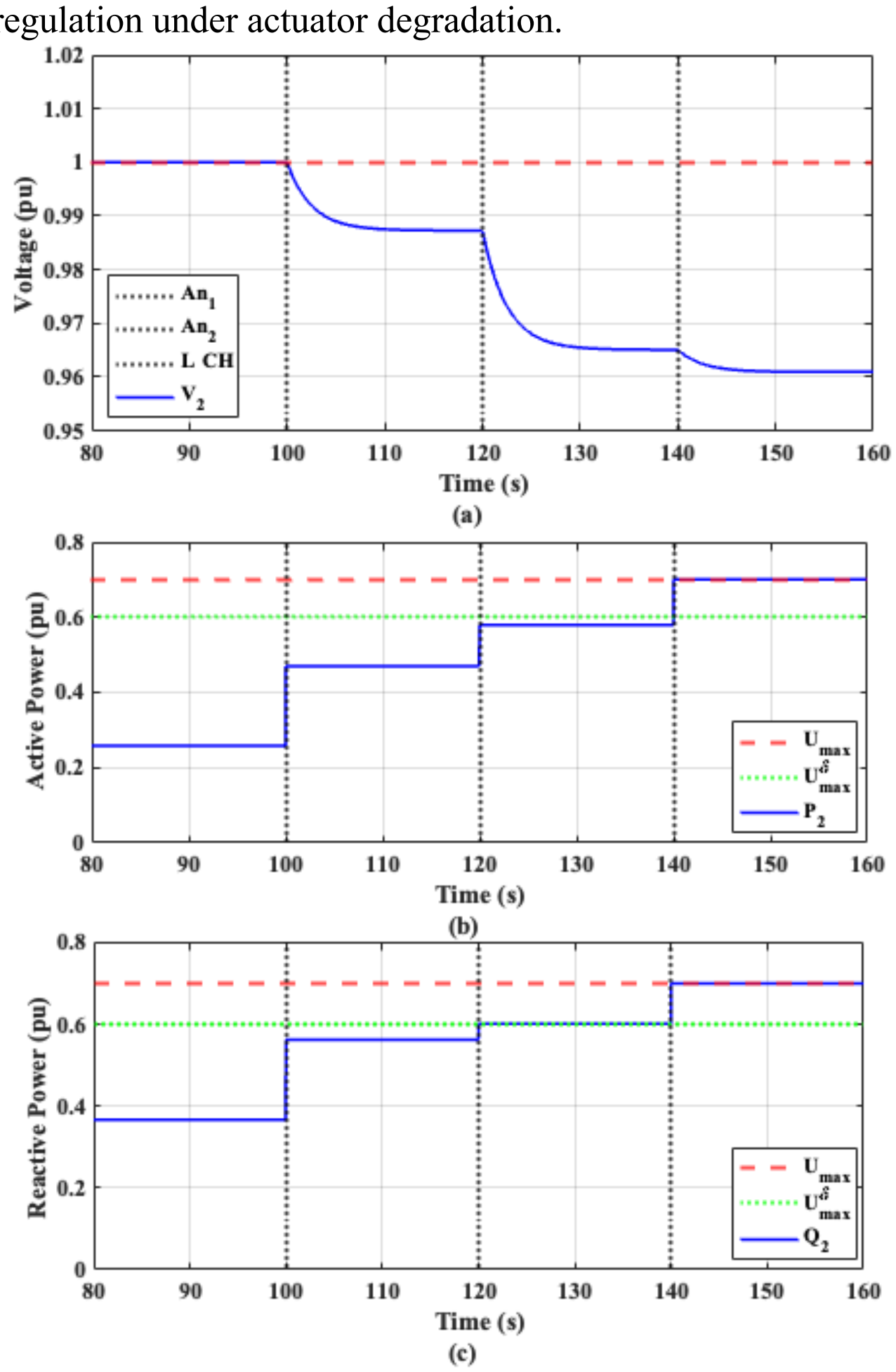

**Fig. 3.** The proposed HOTL controller: (a) Voltage magnitude, (b) Active power, and (c) Reactive power.

*B. Fault Tolerant Controller Response*

The second control framework is a model-based fault-tolerant controller (FTC) [33], whose tracking and control trajectories are shown in Fig. 4. FTC strategies typically rely on a bank of predesigned controller reconfiguration strategies together with an anomaly detection and isolation mechanism. Once a degradation is detected, the controller switches to the closest admissible controller configuration to maintain stability.

As illustrated in Figs. 4(a)–4(c), the FTC maintains acceptable tracking during the first degradation event at t = 100 s, and the voltage trajectories remain stable. However, unlike the proposed HOTL adaptive controller, the FTC exhibits limited adaptability after the second degradation at t = 150 s. As it is shown in Fig.4, the voltage level settles at a lower level, indicating reduced regulation capability under compounded actuator losses. At $t = 200$s, when the load change is introduced. Due to the limited remaining inverter capacity, the controller cannot supply sufficient additional active and reactive power. Consequently, the voltage cannot be recovered to the acceptable operating range and drops below 0.9 pu, reflecting degraded performance and reduced resilience under the combined effects of actuator degradation and load disturbance.

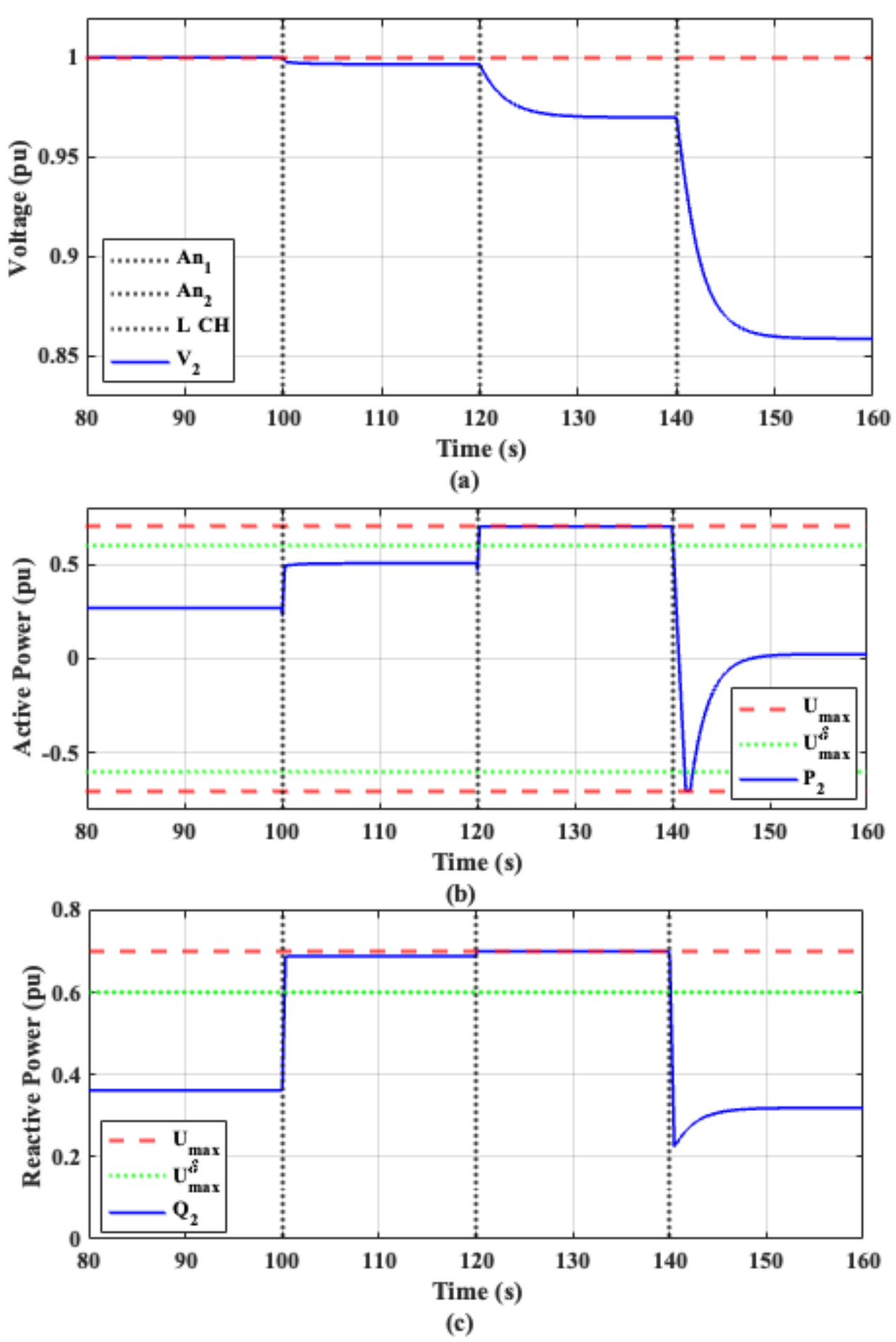

**Fig. 4.** Fault Tolerant Controller: (a) Voltage magnitude, (b) Active power, and (c) Reactive power.

## VI. Conclusion

In this paper, a human-on-the-loop adaptive control architecture was introduced to maintain the resilient operation of IBRs under degradation of inverter capacity, cyber-physical attacks, and uncertain operating conditions. The proposed approach integrates a fast, nonlinear adaptive controller layer with supervisory human input, enabling the control system to incorporate operator situational awareness into the automation process. This architecture overcomes critical challenges that limit traditional fault-tolerant and adaptive control schemes, including the inability to maintain inverter reserve capacity, the absence of structured performance-control mechanisms, and the lack of human-guided resilience management. By introducing two quantitative resilience metrics, GRC and CPD, the method provides a principled framework to balance voltage-tracking performance against the need for preparedness in future grid events. Simulations conducted across various degradation scenarios demonstrated that the proposed controller achieves superior performance by maintaining stable voltage regulation, preventing saturation, and preserving reserve capacity compared to fault-tolerant control method. The results confirm that coordinated human–automation decision-making improves distribution-system resilience.